\documentclass[12pt]{article}

\newcommand{\m}{\mathrm}
\newcommand{\be}{\begin{equation}}
\newcommand{\ee}{\end{equation}}
\newcommand{\ba}{\begin{eqnarray}}
\newcommand{\ea}{\end{eqnarray}}

\usepackage{graphicx}
\usepackage{amssymb}
\usepackage{amsmath}
\usepackage[T1]{fontenc} %for \boldsymbol
\usepackage[ansinew]{inputenc} %for \boldsymbol
\usepackage[nosort]{cite}
\newcommand{\inbar}{\vrule height1.57ex width.4pt depth0pt}
\newcommand{\SW}{\relax{\hbox{$\ \inbar\kern-.285em{\rm S}$}}}

\begin{document}
\thispagestyle{empty}
\begin{center}

\null \vskip-1truecm \vskip2truecm

{\Large{\bf \textsf{How Inflation Might Explain Large-Scale Anisotropy}}}
\vskip0.4truecm

{\large{\bf \textsf{}}}
\vskip0.4truecm

{\Large{\bf \textsf{}}}

\vskip1truecm

{\large \textsf{Brett McInnes}}

\vskip1truecm

\textsf{\\  National
  University of Singapore}

\textsf{email: matmcinn@nus.edu.sg}\\

\end{center}
\vskip1truecm \centerline{\textsf{ABSTRACT}} \baselineskip=15pt
\medskip
Several recent observational analyses have suggested the possible existence of anisotropy in our Universe. It is often thought that this would contradict the inflationary picture. We have argued that this is not the case: if we relax the artificial restriction that spacetime torsion must be zero, then such anisotropies arise naturally, indeed almost inevitably, in a theory of torsional Inflation. This is seen most clearly if one uses a ``$1\,+\,3\,$'' analysis of a torsional spacetime, leading naturally to concepts of intrinsic torsion and extrinsic torsion. An anisotropic quantum fluctuation of the inflaton leads to an extrinsic torsion field which is anisotropic and which grows throughout the inflationary era, even though the intrinsic geometry of the spatial sections is isotropised in the conventional manner. This leads to an anisotropic Hubble field at the \emph{end} of Inflation. The anisotropy takes a very specific and characteristic form, which in principle could be observable on relatively short cosmic scales. On the other hand, an unusual feature of torsional geometry leads to a short interval, \emph{early} in the inflationary era, during which there is a torsion-induced energy flux, which leaves an anisotropic signal in superhorizon modes and thus ultimately at very large scales in the present Universe. Thus in principle torsion can explain anisotropies both on relatively short cosmic scales and also at the largest scales.

\newpage

\addtocounter{section}{1}
\section* {\large{\textsf{1. Inflation vs. Anisotropy }}}
As is well known, the early Universe was very, \emph{but not exactly}, homogeneous and isotropic: there were small fluctuations around a homogeneous, isotropic ``background'' \cite{kn:update}.

The background perfect homogeneity and isotropy is not fully understood \cite{kn:lebowitz,kn:penrose} and will be assumed, not discussed, here: see \cite{kn:spergel} for recent interesting work in this direction. The explanation of the small \emph{deviations} from that state is the office of cosmic \emph{Inflation} \cite{kn:inflationaris,kn:linde,kn:sloth}. Even if the inflaton is initially distributed perfectly uniformly, quantum fluctuations will arise. The enormously rapid expansion produced by the inflationary potential thereafter prevents (in the standard picture\footnote{It is possible to prevent isotropisation by choosing a classically anisotropic inflaton field \cite{kn:soda1,kn:soda2,kn:soda3}, or by exploiting a putative non-trivial spatial topology (see for example \cite{kn:panda,kn:copi}) but we will not consider such possibilities here; we are interested in anisotropy arising spontaneously, not through being ``built-in''.}) inhomogeneous or anisotropic fluctuations from growing inside the horizon \cite{kn:pit}, but those modes which pass outside the horizon ``freeze'': they can be treated as if they were classical stochastic fields (see for example \cite{kn:kiefer}), due to the quantum-to-classical transition associated with decoherence triggered by the existence of the horizon\footnote{The true story here may well be more complex than this sketch suggests: see \cite{kn:bell1,kn:bell2,kn:aurora}. While these recent developments are of great interest, they do not affect the discussion here.}.

When the fluctuations re-enter the horizon, they seed the various structures we see today, and, of course, this inflationary explanation of the origin of those structures (and of the temperature variations in the CMB) has been extremely successful \cite{kn:calabrese} ---$\,$ in fact, many would argue that these successes supersede the original, perhaps somewhat problematic, motivations for Inflation.

In this standard picture, inflationary ``isotropisation'' \cite{kn:pit} implies that there is no apparent way in which cosmic anisotropy can arise or be observable. Recently, however, claims have been made that just such anisotropies, in a wide variety of different systems and at various scales, may have been detected: see for example \cite{kn:eoin0,kn:eoin1,kn:eoin2,kn:eoin3,kn:ani,kn:eoin4,kn:wilt,kn:mig2,kn:maart,kn:boubel,kn:anis,kn:xinwang,kn:aniobserv,kn:salzano,kn:barjou,kn:shubham,kn:souradeep,kn:pilar} (see particularly the survey \cite{kn:eoin}). However, other authors find no evidence \cite{kn:scolnic,kn:stisk,kn:cheng,kn:yongz,kn:sawala}. The latest data suggest \cite{kn:eoinlow} that there may well be anisotropy at relatively short scales, while the situation on the very largest scales remains unclear.

Cosmologically short-scale anisotropies correspond to modes which leave the inflationary horizon late, and thus the underlying matter has been exposed to ``isotropisation'' for a relatively long time: thus they are particularly hard to understand from an inflationary point of view. Nevertheless, we argued in \cite{kn:mcinnes25} that (some) such anisotropies might be explained if we drop the familiar assumption that spacetime \emph{torsion} \cite{kn:blaghehl,kn:lemos,kn:mavro} is negligible, and endow the inflaton with a non-zero spin density which couples to that torsion.

This works because, as was explained in detail in \cite{kn:mcinnes25}, torsion breaks the control which, in General Relativity, the symmetries of the spatial metric impose on the \emph{extrinsic geometry} (which involves both extrinsic curvature and ``extrinsic torsion'') of the distinguished (homogeneous) spatial sections, and it was shown that the extrinsic geometry can be anisotropic even when the intrinsic geometry is isotropised\footnote{There is a formal analogy here with ``latent preferred direction'' cosmologies \cite{kn:latent}.}. That is, extrinsic torsion can easily survive ``isotropisation'', leading naturally to anisotropies at the \emph{end} of Inflation.

However, this argument does \emph{not} apply to anisotropies on the very largest scales, which, if they exist, were seeded very early in the inflationary era. Thus we must begin anew: does torsion give rise to a \emph{different} source of anisotropy, corresponding to modes leaving the inflationary horizon at early times?

Here we will show that such a mechanism does exist. We will show that, in ``torsional Inflation'', an \emph{energy flux} spontaneously arises early in the inflationary era. The extrinsic torsion was still very small at that time, but, due to a peculiarity of torsional geometry (associated with the \emph{torsional Codazzi equation}), the intrinsic torsion ---$\,$ itself perfectly isotropic ---$\,$ amplifies the anisotropy and so permits the flux to be significant. We will see that this energy flux grows briefly to a size that can be bounded by using data at the \emph{end} of Inflation \cite{kn:reheat1,kn:reheat2,kn:reheat3} (where the physics is relatively well understood and (to some extent) constrained \cite{kn:german}) and then rapidly decreases as the intrinsic torsion ``inflates away''.

In other words, torsion naturally and (almost) inevitably implies that there is indeed a simple way of perturbing the ``spin-bearing inflaton'' such that the anisotropic perturbation grows, in the \emph{early} inflationary era, to a magnitude which can be bounded, and which is then suppressed, so that only large scales are affected. In short, large-scale anisotropy, if it is eventually shown to exist, will \emph{not} necessarily undermine Inflation. On the contrary, if we are fortunate, it could turn out to be another successful inflationary prediction.

In order to avoid confusion, we will focus (after the next, motivational Section) on the behaviour of the intrinsic and extrinsic \emph{torsions}. We will assume throughout this work that the purely metric geometry is always as symmetric as possible. This means in particular that the spacetime metric geometry is as in the usual FRW cosmologies, with a well-defined cosmic scale factor $a(t)$. Anisotropies only arise in the \emph{extrinsic} geometry, as explained above. In reality, of course, the ``isotropisation'' of the metric geometry occurs simultaneously \cite{kn:pit}, but we leave this to one side for the sake of simplicity.

We begin with a brief discussion of the geometric meaning of standard Inflation; that prepares the way for a parallel discussion for torsional Inflation.

\section* {\large{\textsf{2. What Inflation Does to Spatial Curvature }}}
In this Section (only) we will set the spacetime torsion tensor to zero: the purpose here is to motivate what we subsequently do with torsion.

We will suppose that the distinguished spatial sections normally assumed to exist in cosmology have both intrinsic and extrinsic geometries which become more and more symmetric as we examine them farther back in time: see \cite{kn:lebowitz,kn:penrose}. Our intention here is not just to accommodate the fact that the ``gravitational entropy of the earliest Universe'' must have been extremely small, but also, as we stated earlier, to avoid building in anisotropies right from the beginning: we want to explain anisotropy, not assume it.

This could involve taking a limit to approach the highly symmetric but \emph{singular} state which (neglecting quantum-gravitational effects, which we always do here) probably \cite{kn:ghazal} preceded Inflation; or it could entail a direct examination of the highly symmetric non-singular initial state assumed in ``creation from nothing'' scenarios\footnote{There is of course a vast literature on this idea, and on the various difficulties one encounters when attempting to reconcile it with Inflation: see \cite{kn:juan} for a summary, and for suggestions as to how those difficulties might be dealt with. See also \cite{kn:wine} for remarkable recent developments.} \cite{kn:lehners}. As this latter case is simpler, let us discuss it first; we will return to the singular case later.

In ``creation from nothing'' theory, the (Lorentzian) Universe comes into existence along a small spacelike hypersurface of \emph{zero extrinsic curvature}. This is necessary for a smooth transition from a ``Euclidean'' (positive signature) domain.

As we discussed earlier, the principal role of Inflation is to give rise to perturbations around this initial state. But Inflation also, of course, controls the classical evolution of the spacetime. In this Section, we will be concerned with that role for the inflaton.

Since the initial energy density is large and positive, then, from the usual Friedmann equation and the vanishing of the extrinsic curvature, the intrinsic curvature of the initial hypersurface must have a large positive value\footnote{Arguments in favour of the positivity of the spatial curvature (when the torsion is zero) at early times have recently been advanced in \cite{kn:damien1,kn:damien2,kn:damien3}; here, we apply this to creation-from-nothing cosmologies, and not to the (however, closely related) bouncing cosmologies considered there. We note in this connection that \emph{some} torsional cosmologies bounce; but some do not, and that includes the specific models to be considered below, which involve torsion components unlike those usually studied.}: it is a three-sphere with extremely large positive scalar curvature.

This, then, is the ``$1\,+\,3\,$ geometry'' of the beginning of Inflation in such theories: enormous intrinsic curvature, but zero extrinsic curvature.

This is very remarkable, because the geometry of the spatial sections we currently inhabit has the diametrically \emph{opposite} form: we see spacelike sections with very small intrinsic curvature \cite{kn:sunny2,kn:sunny3,kn:efstath,kn:sunny1,kn:sunny4,kn:eleonora,kn:camphuis} but relatively large extrinsic curvature (measured by the Hubble parameter ---$\,$ see below). Somehow, the spatial curvature has been converted from intrinsic to extrinsic form.

In principle at least, Inflation explains this peculiar transformation very neatly. The spacetime geometry is, to a good approximation, that of the version of de Sitter spacetime foliated by three-spheres with intrinsic curvature given by
\begin{equation}\label{INTRINSIC}
R(W,\,Z,\,X,\,Y)\,=\, L^{-2}\m{sech}^2(ct/L)\left[g(Y,\,Z)\,g(W,\,X)\;-\;g(X,\,Z)\,g(W,\,Y)\right],
\end{equation}
where $W,\,Z,\,X,\,Y$ are spacelike tangent vectors, $t$ represents proper time, and where $L$ is the length scale defined by the inflaton energy density.

Of course, the intrinsic curvature decays, ultimately exponentially rapidly, as time passes. But this forces the \emph{extrinsic} curvature to \emph{grow}. For de Sitter spacetime has constant curvature. (In physical language: the inflaton itself does not ``inflate away''; in exact de Sitter spacetime, the total energy density is a constant.) So the Gauss formula (\cite{kn:kobnom2}, Chapter VII), which expresses the total curvature as the sum of the intrinsic and extrinsic curvatures of a spatial section, implies that any loss of intrinsic curvature must be exactly compensated by an increase in the extrinsic curvature.

And indeed, in the same notation, the extrinsic curvature grows in such a manner that the total coefficient is (from an elementary hyperbolic geometry identity) constant:
\begin{equation}\label{EXTRINSIC}
R^E(W,\,Z,\,X,\,Y)\,=\, L^{-2}\tanh^2(ct/L)\left[g(Y,\,Z)\,g(W,\,X)\;-\;g(X,\,Z)\,g(W,\,Y)\right].
\end{equation}
To summarise: the intrinsic curvature ``inflates away'', but this is exactly compensated by the ``inflating up'' of the extrinsic curvature\footnote{In an inflationary spacetime which is not exactly de Sitter, the compensation might not be exact; but the point is that there must be \emph{some} compensation: the extrinsic curvature cannot always vanish.}. Note that the coefficient on the right side of equation (\ref{EXTRINSIC}) is (up to a factor of $c^2$) \emph{the square of the Hubble parameter}, so the large value of the latter at the end of Inflation (and therefore the expansion we observe now) is precisely a manifestation of this growth of the extrinsic curvature.

From a ``$1\,+\,3\,$'' point of view, then, \emph{Inflation is a mechanism for converting intrinsic spatial curvature to extrinsic curvature}\footnote{The reader will be aware that most discussions of Inflation assume flat spatial sections from the outset ---$\,$ that is, they make use of the subspace of de Sitter spacetime that can be foliated by flat spatial sections. We regard this as a limiting case of the above discussion: the intrinsic curvature in this limit is ``already'' zero at the outset, so the extrinsic curvature is ``already'' equal to the full spacetime curvature, and it must therefore be constant, as indeed it is (the Hubble parameter is constant in this version of de Sitter spacetime). We will see another version of this in the torsional case, below.}.

It will be useful to phrase this mechanism in more general terms.

In standard General Relativity, initial conditions take the form of specifying the spatial metric $g$ on a spacelike hypersurface $\Sigma,$ along with a $(0,\,2)$ tensor $h$; the two are related to each other and to the energy density\footnote{We always absorb the cosmological constant into this quantity.} $\rho$ by one of the constraint equations:
\begin{equation}\label{A}
{16\,\pi \,G\over c^4}\,\rho\,=\,\m{Scal}_g \,+\,\left(\m{Tr}\,\hat{h}\right)^2\,-\,\m{Tr}\,\hat{h}^2,
\end{equation}
where $G$ and $c$ are as usual, $\m{Scal}_g$ is the three-dimensional scalar curvature associated with $g$, $\hat{h}$ is the $(1,\,1)$ version of $h$, and $\m{Tr}$ denotes the trace map. When the spatial sections are perfectly isotropic around an arbitrarily chosen point, Schur's lemma implies that $h = Hg/c,$ where $H$ is the Hubble parameter, and then one shows easily that the usual Friedmann equation is just a special case of equation (\ref{A}).

As the system evolves into a foliation of the spacetime, one finds that $h$ is the \emph{second fundamental form} of that foliation: it describes the extrinsic geometry of the spatial sections, the manner in which those sections bend into spacetime. When $h = Hg/c,$ one sees that the expansion of the Universe, quantified by $H$, is simply an aspect of the extrinsic geometry of those sections.

The energy density of the inflaton is approximately constant throughout Inflation, and so equation (\ref{A}) generalises our earlier observations about the geometric interpretation of Inflation: if the intrinsic geometry, represented by $\m{Scal}_g$, is not a constant, then the extrinsic geometry, represented by the terms involving $h$, has to compensate. In particular, if $\m{Scal}_g$ tends to zero, as it typically will in the Inflationary context, then the total of the remaining terms involving the second fundamental form \emph{must} increase.

Small quantum fluctuations will give rise to inhomogeneities in $g$, but these are held in check by ``isotropisation'' \cite{kn:pit} inside the horizon. It turns out that, \emph{when the spacetime torsion is zero} \cite{kn:mcinnes25}, the symmetries of $h$ are the same as those of $g$, so the extrinsic geometry is also isotropised in this case. Equation (\ref{A}) then tells us that, apart from the well-known anisotropies due to quantum fluctuations as discussed earlier, the matter field too must be almost exactly isotropic, like $g$ (and therefore $\m{Scal}_g$) and $h$, throughout Inflation.

Let us see what happens when torsion is present.

\addtocounter{section}{1}
\section* {\large{\textsf{3. What Inflation Does to Spatial Torsion }}}
We assume that the inflaton carries both an energy density and a spin density. (For earlier work on ``spinor Inflation'', see \cite{kn:spinor} and papers citing it.) The inflaton does not ``inflate away'', so these densities must, likewise, not decay, at least not to zero.

We pause here to remark that we are not attempting to construct a complete ``torsional cosmology'' here, or even a complete theory of torsional Inflation. Our objective is more modest: to understand in detail how torsional Inflation can naturally and inevitably give rise to large-scale anisotropy. For much more complete discussions of torsional cosmology, see for example \cite{kn:orazi1,kn:orazi2,kn:racioppi}, and particularly \cite{kn:izau}.

In view of this, we will assume that, just as the energy density of the inflaton is approximately constant for most of the inflationary era, so too is the spin density. In broad terms, one could (for example) model the inflaton as a pseudoscalar fermionic condensate in a theory with a self-interaction potential, with a classical spinor field $\psi$ such that the expectation value of $\bar{\psi}\gamma^5\psi$ is constant. Again, we stress that there may well be many possibilities of this kind, and that we do not insist that, in reality, the spin density should be \emph{exactly} constant.

We will be using the Einstein-Cartan theory (and not one of its many generalizations). This is strictly for reasons of simplicity: in this theory, it is possible to carry out all of our analyses exactly. This brings out our main conclusions very clearly, and we anticipate that these conclusions apply, in a qualitative way, to more general torsional theories (in which, for example, torsion is able to propagate). One hopes that it will be possible to obtain more satisfactory quantitative results using those theories. (See in this connection \cite{kn:mud}, which addresses several issues directly relevant to our discussion here.)

In this theory, the spin current tensor $S^*$ is related to the spacetime torsion $T^*$ in a simple manner\footnote{The asterisks remind us that we are speaking of \emph{spacetime} quantities. Symbols without asterisks refer to purely spatial quantities.}. In the cases of interest in this work (where the torsion will always be traceless, as we will discuss below), the relation is particularly simple: we have
\begin{equation}\label{EC}
T^*\,=\,{8\,\pi \,G\over c^4}S^*,
\end{equation}
where of course the coefficient is the usual one in General Relativity. Up to this constant, the full spacetime torsion is essentially the spin current tensor in this theory.

When it is evaluated on vectors tangential to $\Sigma$, the spacetime torsion for a given foliation by spacelike hypersurfaces $\Sigma$ splits, in a manner precisely analogous to the familiar splitting of the curvature (as discussed above) in semi-Riemannian geometry: we have \cite{kn:mci}
\begin{equation}\label{B}
T^*(X,\,Y)\,=\,T(X,\,Y)\,+\,T^E(X,\,Y),
\end{equation}
where $X$ and $Y$ are tangent vectors to $\Sigma$, $T^*$ is the spacetime torsion as above, $T$ is the torsion tensor of the induced connection on $\Sigma$ (that is, it represents the \emph{intrinsic torsion}), and $T^E$ is the \emph{extrinsic torsion}, given by
\begin{equation}\label{EXTORH}
T^E(X,\,Y)\,=\,\left[h(X,\,Y)\,-\,h(Y,\,X)\right]\xi \,=\, \varkappa(X,\,Y)\,\xi,
\end{equation}
where $h$ is, as before, the $(0,\,2)$ version of the second fundamental form, where $\xi$ is the future-pointing unit normal field to $\Sigma$, and where $\varkappa$ is (twice) the antisymmetric part of $h$. We shall call $\varkappa$ the \emph{extrinsic torsion form}. (It is closely related to the \emph{contortion tensor} in torsional geometry.)

The extrinsic torsion is the analogue of the extrinsic curvature, in the sense that both are determined by the manner in which $\Sigma$ is embedded: that is, both are determined by the second fundamental form. Note that $h$ is symmetric if and only if the extrinsic torsion is zero.

Our objective is to show that, during torsional Inflation, the torsion behaves in precisely the same manner as the curvature: that is, Inflation should convert intrinsic torsion to extrinsic torsion.

In view of this, we will assume that the intrinsic and extrinsic torsions defined by the distinguished (homogeneous) foliation are the only non-zero spacetime torsion components. In general, this is not so \cite{kn:mcinnes25}, but it turns out that the inclusion of these additional torsion components would not materially change our conclusions; it would only complicate the analysis unnecessarily. The geometric interpretation of this assumption is that, in torsional Inflation, the expansion of the Universe is completely controlled by the non-trivial extrinsic geometry of the distinguished spatial sections, exactly as in the zero-torsion case\footnote{Specifically: we are excluding the torsion components corresponding to the $(1,\,1)$ tensor $T^*(\xi,\,\_)$; setting this tensor to zero implies that the \emph{Hubble tensor} \cite{kn:mcinnes25} (see below), describing a possibly anisotropic expansion, coincides with the second fundamental form, as it does in the zero-torsion case. The inclusion of this ``bulk torsion'' would only affect our results if we artificially forced the extrinsic torsion to be identically zero, as is done in those torsional cosmologies where isotropy is rigorously enforced from the outset; see for example \cite{kn:barrow}. This is of course the diametric opposite of what we want here.}.

Let us attempt to construct a precise torsional analogue of our discussion of the way Inflation works in the zero-torsion case discussed above.

Unlike the de Sitter spacetime we discussed earlier, the spacetime metric in our case proves to be singular, so we cannot assume the existence of an \emph{initial} hypersurface: instead, we more cautiously assume that, as we proceed back in time, the distinguished hypersurfaces become ever more symmetric and have a large intrinsic torsion but an arbitrarily small extrinsic torsion.

In practice, this means that we will truncate the geometry at some value $a_0$ of the cosmic scale factor ---$\,$ recall that the metric geometry is of FRW form here ---$\,$ and only discuss the situation subsequent to the corresponding proper time, $t_0$; the spatial section at that time plays the role of the ``initial'' hypersurface. The reader can think of $t_0$ as \emph{the time when Inflation began}. (It will turn out that $t_0$ and $a_0$ are fixed by physical parameters, but their actual numerical values will not be important for our conclusions.)

Now just as the unit three-sphere $S^3$ has a canonical (maximally symmetric) Riemannian metric, so also does it have a canonical (maximally symmetric) torsion tensor: the torsion of the connection naturally defined by the fact that $S^3$ is parallelizable\footnote{Meaning that there is a global section of the bundle of orthonormal frames.}, which in turn is related to the fact that $S^3$ is the group manifold for $SU(2)$. This connection is a metric connection with respect to the Riemannian metric defined by the Cartan-Killing bilinear form. (See \cite{kn:kobnom2}, Chapter X. Notice that all of the FRW spacetimes have parallelizable spatial sections, so all fit very naturally into any theory of spacetime torsion.)

In standard Inflation, if we choose a spacetime modelled on global de Sitter spacetime, the metric on the spatial sections is always a multiple of the metric on the unit three-sphere; this ensures that the spacetime is always spatially isotropic around every point. We shall assume that, likewise, the intrinsic spatial torsion is always a multiple of the canonical torsion tensor on the unit three-sphere, so that the intrinsic torsion is always isotropic around each point.

For the intrinsic torsion to be isotropic, its $(0,\,3)$ version $\emph{\text{\v{T}}}$ must be completely antisymmetric\footnote{This fact, combined with equation (\ref{EXTORH}) and the assumption that the intrinsic and extrinsic torsions are the only non-zero components of the spacetime torsion, imply that $T^*$ is traceless, and that is why equation (\ref{EC}) takes its particularly simple form.}. Now note that the connection defined by the parallelization is flat; its curvature tensor, and therefore its scalar curvature, must be zero. Expressing that scalar curvature in terms of the Riemannian scalar curvature Scal$_g$ of the three-sphere, one finds, using the complete antisymmetry of $\emph{\text{\v{T}}}$, that the vanishing of the scalar curvature implies
\begin{equation}\label{C}
T^2 \,=\,4\,\m{Scal_g},
\end{equation}
where the square of $T$ is defined, using the spatial metric, as $T_{ijk}T^{ijk}$. Since we anticipate ---$\,$ of course, we still have to demonstrate this ---$\,$ that Inflation reduces Scal$_g$ towards zero very rapidly, clearly the same is true of the square of the intrinsic torsion (with respect to the positive-definite spatial metric) and hence the same is true also of the intrinsic torsion itself.

Under our assumptions here, the field equation (\ref{EC}) implies that the sum (in a sense we will make precise) of the intrinsic and extrinsic torsions must be constant with time. It follows that, in torsional Inflation, the magnitude of the extrinsic torsion \emph{must} grow through the inflationary era. This is of course entirely analogous to the fate of the extrinsic curvature, in the zero-torsion case discussed earlier: both extrinsic objects have to grow, to compensate for the decay of their intrinsic counterparts.

The (Riemannian) scalar curvature of the three-sphere is $6/R^2$ if the radius is $R$, so we have from equation (\ref{C}) that $T^2 \,=\,24/R^2$. On the other hand, the $(0,\, 3)$ version of $T$, which we called $\emph{\text{\v{T}}}$ earlier, is a multiple of the volume form on the three-sphere, $\Omega,$ which has square equal to 6. It follows that $\emph{\text{\v{T}}}\,=\,{2\over R}\,\Omega.$ (This can be confirmed by explicitly constructing the torsion in terms of the left-invariant vector fields on $SU(2)$.) We note in passing that this equation only makes sense because space is three-dimensional.

We are assuming that the spacetime metric during Inflation was of FRW form with a typical three-spherical spatial section $\Sigma$ (with volume form $\Omega_{\Sigma}$) and with scale factor $a(t)$, where $t$ is proper time. Then from the above discussion we have in our case
\begin{equation}\label{D}
\emph{\text{\v{T}}}\,=\,{2\over a}\,\Omega_{\Sigma}.
\end{equation}
Let $\sigma$ be the spin density of the inflaton: as explained earlier, for simplicity we assume that it, like the energy density, should be (to a good approximation until Inflation begins the transition to reheating \cite{kn:reheat1,kn:reheat2,kn:reheat3}) constant. Since the extrinsic torsion is assumed to be zero at the ``initial'' time $t = t_0$ (when $a = a_0$), the full torsion at that time is approximately equal to the intrinsic torsion. Thus $\sigma$ is a multiple of the coefficient of the volume form in equation (\ref{D}): equations (\ref{EC}) and (\ref{D}) give us (taking into account a factor of $c$ to distinguish the spin current from the spin density)
\begin{equation}\label{E}
\sigma\,=\,{c^3\over 4\pi G a_0}\,.
\end{equation}

From equation (\ref{EXTORH}) we know that the extrinsic torsion is described by a two-form, $\varkappa.$ \emph{But any such object is intrinsically anisotropic}: a non-zero two-form on an odd-dimensional real manifold cannot be isotropic (see Appendix 5 of \cite{kn:kobnom1}). This geometric anisotropy presumably begins with a quantum fluctuation (which determines a distinguished direction in a probabilistic manner), and thereafter it grows, as we have seen. We stress that this anisotropy continues to grow throughout the inflationary era: it is forced to do so by the continuing dilution of the intrinsic torsion (equation (\ref{D})). It is therefore distinct from the usual anisotropies associated with standard Inflation.

Thus we conclude that \emph{torsional Inflation inevitably induces (a specific kind of) anisotropy}. The initially ultra-symmetric, low-entropy spatial geometry has proved to be ``unstable'': a tiny fluctuation has triggered a rapidly growing, eventually classical, loss of symmetry.

Let us examine this type of anisotropy in more detail.

The matrix of $\varkappa$, if it is not zero, has one eigenvector with zero eigenvalue, and two other eigenvalues which are pure imaginary and mutually conjugate; there is a single real eigenvector, with zero eigenvalue, \emph{defining a distinguished direction}. There exists a global field of orthonormal bases (which we take to be one that parallelizes\footnote{Each vector field in such a global field of bases has zero spatial covariant derivative with respect to all of the others, so that the connection coefficients vanish for this basis.} $S^3$), with the zero-eigenvalue eigenvector as the third member, with respect to which the matrix of $\varkappa$ is
\begin{equation}\label{F}
[\varkappa]\,=\, \kappa\left(\begin{array}{ccc}
  0 & 1 &0  \\

 - 1 & 0 & 0 \\
 0 & 0 & 0 \\
  \end{array}\right),
\end{equation}
where $\kappa$ is a real quantity. According to our assumption that the spatial geometry (intrinsic and extrinsic) should be as symmetric as possible, $\kappa$ can depend on time but not on spatial position. (We could allow it to depend on position, but, once again, we do not want to build anisotropy in from the outset, to the extent that this can be avoided. If we had allowed this, then there would have been a strong anisotropy at the \emph{end} of Inflation ---$\,$ and not just in the spatial geometry, but also in the energy density, see equation (\ref{I}) below.)

Taking the Hodge dual of $\varkappa$, we obtain a three-vector parallel to the distinguished direction, with component precisely $\kappa$ in that direction, so $\kappa$ is the scalar function of time that parametrises the extrinsic torsion.

It turns out \cite{kn:mcinnes25} that $\kappa$ also directly parametrises the anisotropy of the vector field describing the rates of change of positions of (fictitious) point particles imagined to be embedded in the inflaton, pictured as a ``fluid''. This field is obtained by letting a certain tensor, the \emph{Hubble tensor}, act on the position vector; this is of course the generalisation of the Hubble-Lema\^{i}tre Law to the anisotropic case. The Hubble tensor factorises, in a polar decomposition, into a product of a certain rotation transformation with a transformation $\eta$ which has a matrix, relative to the above basis, taking the form
\begin{equation}\label{FF}
[\eta]\,=\, \left(\begin{array}{ccc}
  \sqrt{H^2 + \kappa^2/4} & 0 &0  \\
 0 & \sqrt{H^2 + \kappa^2/4} & 0 \\
 0 & 0 & H \\
  \end{array}\right),
\end{equation}
where $H$ is the usual Hubble parameter. Evidently $H$ describes the rate of expansion (during Inflation) in the distinguished direction, but there is a different rate in the perpendicular directions, and the difference is controlled by $\kappa$.

The rotational factor in the polar decomposition of the Hubble tensor takes the form
\begin{equation}\label{WWW}
[Q]\,=\, \left(\begin{array}{ccc}
  {H\over \sqrt{H^2 + \kappa^2/4}} &  {- \kappa/2\over \sqrt{H^2 + \kappa^2/4}}  &0  \\

 {\kappa/2\over \sqrt{H^2 + \kappa^2/4}} & {H\over \sqrt{H^2 + \kappa^2/4}} & 0 \\
 0 & 0 & 1 \\
  \end{array}\right).
\end{equation}
The rotation is around the same axis as the distinguished axis discussed above, and the angle of the rotation is again determined by $\kappa$.

Thus we see that the extrinsic torsion leaves a very specific and characteristic pattern in the Hubble field at the time of reheating (the end of Inflation), controlled by a single parameter, the value of $\kappa$ at that time. We stress that this phenomenon is indeed most relevant at the \emph{end} of Inflation: this is simply because the extrinsic torsion grows throughout Inflation, so it is small at early times. Therefore this kind of anisotropy is related to inflationary modes that leave the horizon very late, and consequently it is relevant to fairly small-scale anisotropies. \emph{It cannot explain anisotropies on the largest scales}.

The spin density $\sigma$ is, by equation (\ref{EC}), just a multiple of the relevant torsion component. It therefore has, at any time, two parts: one (equation (\ref{D})) contributed by the intrinsic torsion, and the other, described by $\kappa,$ contributed by the extrinsic torsion, as explained above. Since we require $\sigma$ to be at least approximately constant, we have now, from equation (\ref{D}),
\begin{equation}\label{G}
\kappa \,=\,2 \left({1\over a_0}\;-\,{1\over a}\right);
\end{equation}
as explained above, we restrict to the domain $a(t) \geq a_0$.

Using equation (\ref{E}) we can express this as
\begin{equation}\label{H}
\kappa \,=\,2 \left({4\pi G \sigma\over c^3}\;-\,{1\over a}\right).
\end{equation}

Now we need the torsional analogue of the Friedmann equation (in the specific case of the Einstein-Cartan theory). As in the case of General Relativity, this equation is really nothing more than one of the \emph{constraints} which have to be satisfied by the ``initial'' data \cite{kn:luz}. It was given in \cite{kn:mci}, to which we refer the reader:
\begin{equation}\label{I}
{16\pi G\,\rho\over c^4}\,=\,\m{Scal}\,+\,\left(\m{Tr}\,\hat{h}\right)^2\,-\,\m{Tr}\,\left(\hat{h}^T\hat{h}\right);
\end{equation}
here, as before, $\rho$ is the (constant) energy density, but now $\m{Scal}$ is the scalar curvature of the full linear connection on the spatial section. Note that, in the presence of (extrinsic) torsion, $h$ is not a symmetric bilinear form, so $\hat{h}^T\hat{h}$ is not the square of $\hat{h}$; this proves to be important (see \cite{kn:mci}).

As we have discussed, in our case the spatial sections are flat with respect to the torsional connection, though of course not with respect to the metric. We therefore have $\m{Scal} = 0,$ and so the right side of equation (\ref{I}) depends only on the second fundamental form. This is precisely as in the case of ordinary de Sitter spacetime foliated by flat (in the usual sense) spatial sections, as discussed earlier.

The difference is that the second fundamental form is no longer given only by the Hubble parameter. As is discussed at length in \cite{kn:mcinnes25}, if we assume that the spatial metric is exactly isotropic around each point (that is, if we leave aside the anisotropies due to quantum fluctuations), then (with our standing assumption that the ``bulk torsion'' mentioned earlier is absent) $h$ takes the form
\begin{equation}\label{J}
h\,=\,{H g\over c} \,+\,{1\over 2}\,\varkappa,
\end{equation}
where $H$ is the usual Hubble parameter, $H = \dot{a}/a,$ where the dot denotes a derivative with respect to proper time.

Substituting this into equation (\ref{I}), one finds, using equation (\ref{F}), that the generalised Friedmann equation takes the form
\begin{equation}\label{K}
{8\pi G\,\rho\over c^4}\,=\,{3\,H^2\over c^2}\,-\,{1\over 4}\,\kappa^2,
\end{equation}
showing explicitly that the extrinsic torsion affects the scale factor. Notice that this means that it is not entirely clear that the scale factor will evolve in the manner we require if this theory is to be ``inflationary'': we have to confirm that. (Notice too that, while the extrinsic torsion form $\varkappa$ is not isotropic, the scalar $\kappa$ \emph{is}, under our assumptions, and consequently so are $\rho$ and $\sigma$. The anisotropy is in the Hubble field, not in the densities.)

Using equation (\ref{H}), we can express equation (\ref{K}) as
\begin{equation}\label{L}
{8\pi G\,\rho\over c^4}\,=\,{3\,H^2\over c^2}\,-\,\left({4\pi G \sigma\over c^3}\;-\,{1\over a}\right)^2.
\end{equation}
Because $\rho$ and $\sigma$ are approximately constant until reheating sets in, this equation has a simple exact solution, which can be stated as follows. First, define length scales $L_{\rho}$ and $L_{\sigma}$ by
\begin{equation}\label{M}
{1\over L_{\rho}^2}\,=\,{8\pi G\,\rho\over 3 c^4}, \;\;\;\; {1\over L_{\sigma}}\,=\,{4\pi G\,\sigma\over \sqrt{3} c^3};
\end{equation}
since both are lengths, the ratio $L_{\sigma}/L_{\rho}$ gives us a dimensionless (inverse) measure of the relative size of the spin density to the energy density. Next, define an overall length scale $\ell$ by
\begin{equation}\label{N}
{1\over \ell}\,=\,\sqrt{{1\over L_{\rho}^2}\,+\,{1\over  L_{\sigma}^2}}.
\end{equation}
In the absence of spin or torsion, $\ell = L_{\rho}$ is just the usual de Sitter length scale. Notice that we always have $\ell \leq L_{\rho}$ and $\ell \leq L_{\sigma}$.

Then the solution for the scale factor is given by
\begin{equation}\label{O}
a(t)\,=\,{\ell \over \sqrt{3}}\left(\sinh(ct/\ell)\,-\,{\cosh(ct/\ell)\,-\,1\over \sqrt{1 + {L_{\sigma}^2\over L_{\rho}^2}}}\right),
\end{equation}
the constant of integration being fixed by the condition $a(0) = 0.$

The function $a(t)$ is always monotonically increasing, and clearly at late times the growth is effectively exponential; so we have the usual explosive inflationary expansion: ``torsional Inflation'' is indeed a kind of Inflation.

Our earlier anticipation that the scalar curvature Scal$_g$ defined by the metric tends to zero as time passes is also thus confirmed: and so, from equation (\ref{C}), is our prediction that the intrinsic torsion decays to zero as Inflation proceeds.

At $t = 0$ there is a genuine singularity: for example, the Kretschmann scalar (computed from the metric tensor only) diverges there. That is, the timelike curves representing the trajectories of free particles\footnote{In torsional theories these curves are typically not geodesics of the full linear connection. In \cite{kn:mcinnes25} we explained this in terms of the close relation between anisotropy and torsion, precisely the relation we are exploiting here. (The point is that there is no reason to expect free particles to have straight worldlines when spacetime itself is not isotropic around every point.) Instead, the trajectories are curves of extremal proper time, and these of course are controlled by the metric aspect of the geometry, in particular, by the metric Kretschmann scalar. Note in this connection that photon trajectories are, for the same reason, not directly affected by torsion, and so we do not expect the effects discussed in this work to have any \emph{direct} consequences for the CMB (though they might have indirect effects because they can affect large-scale structures).} cannot be extended arbitrarily far to the past. As mentioned earlier, this is to be expected generically \cite{kn:ghazal} in inflationary theories which do not take quantum-gravitational effects into account. In any case, in the present work, we explicitly confine attention to values of the scale factor above some minimum value, so the singularity does not directly concern us.

In summary, then: we have a torsional spacetime geometry in which the intrinsic torsion, the extrinsic torsion, and the spatial metric vary with time in known ways: see equations (\ref{D}), (\ref{H}), and (\ref{O}). The extrinsic torsion is inherently anisotropic (equation (\ref{FF})), and $\kappa,$ the function quantifying the anisotropy, has to grow through the inflationary era, approaching an upper bound given (equation (\ref{H})) by
\begin{equation}\label{OO}
\kappa_{\m{max}}\,=\,{8\pi G \sigma \over c^3}.
\end{equation}
Evaluating equation (\ref{K}) at reheating (indicated by the Re subscript), we therefore have
\begin{equation}\label{OOO}
{H_{\m{Re}}^2\over c^2}\,=\, {8\pi G\,\rho\over 3 c^4}\,+\,{16\pi^2G^2\sigma^2\over 3c^6};
\end{equation}
here the densities are evaluated at reheating, which means that $\rho$ is related to the energy density of the radiation produced by the decay of the inflaton.

Now equations (\ref{M}) and (\ref{N}) allow us to write this as simply
\begin{equation}\label{OOOO}
{H_{\m{Re}}\over c}\,=\,{1\over \ell}.
\end{equation}
The expression for the scale factor (equation (\ref{O})) shows that the quantity $\ell/c$ (defined in equation (\ref{N})) is the natural unit of time here. Of course, $a(t)$ is not exactly exponential at early times, but it very rapidly becomes indistinguishable from it, so $\ell/c$  corresponds to what we might call ``late'' e-folds\footnote{In fact, the rate of expansion is close to exponential even at early times, so these late e-folds do not differ greatly at any time from what we normally think of as e-folds, which in practice are the conventional measure of ``time'' in inflationary theory. (Notice however that, since $\ell \leq L_{\rho}$, the presence of torsion does marginally hasten the expansion in the later part of Inflation, as measured by \emph{proper} time.)}.

On the other hand, in inflationary theory the value of the Hubble parameter when reheating is complete is determined by the decay rate of the Inflaton \cite{kn:reheat1}. We therefore regard $H_{\m{Re}}$, and thus (from equation (\ref{OOOO})) $\ell$, as a fixed quantity, in principle to be computed from the details of the specific inflationary model we choose. From equation (\ref{N}) we see that fixing $\ell$ leaves us with one free parameter, which we take to be the dimensionless quantity $L_{\sigma}/L_{\rho}$.

Equation (\ref{OOO}) informs us that there are two contributions to the Hubble parameter at reheating: one associated with the radiation energy density, and one due to the spin density, or, equivalently (since by then the intrinsic torsion has inflated away), to the presence of the extrinsic torsion. Since we are assuming that the total is fixed, this means that the radiation density is \emph{reduced} by the presence of the extrinsic torsion, and so is the temperature. But this \emph{reheating temperature} is, at least in principle, constrained to some extent by its relations with other physical variables and phenomena \cite{kn:german}. This means, again in theory, that the magnitude of the extrinsic torsion at the end of Inflation can also be constrained.

To take a concrete example, the universe has to be radiation-dominated and in thermal equilibrium before Big Bang Nucleosynthesis begins, so we must not allow the reheating temperature to fall below the BBN temperature. Equation (\ref{OOO}) therefore imposes a hard upper bound on the size of the spin density relative to the energy density, that is, \emph{a hard lower bound on $L_{\sigma}/L_{\rho}$}.

Unfortunately, while the lower bound imposed in this manner is indeed firm, that does not mean that we can specify it precisely. This is so because the numerical constraints on the reheating temperature are currently very weak and very model-dependent. Fortunately, as we will see later, there are other reasons to think that $L_{\sigma}/L_{\rho}$ cannot be small, in the sense of being significantly smaller than unity.

Obtaining an upper bound on $L_{\sigma}/L_{\rho}$ is still more difficult; again, the problem is the lack of precise data on the reheating temperature. But this could change in the foreseeable future: for example, a detection of stochastic gravitational waves from the reheating era might well change the situation drastically. We will discuss $L_{\sigma}/L_{\rho}$ further in the next Section.

In summary, the presence of extrinsic torsion towards the end of Inflation does have physical consequences: it forces the Hubble field at that time to be anisotropic, and this affects the ``initial conditions'' for the subsequent evolution. But precisely because all this pertains to the final era of Inflation, this kind of anisotropy affects scales which, by cosmological standards, are small. What, then, of observations suggesting anisotropy at very large scales? We now turn to this.

\addtocounter{section}{1}
\section* {\large{\textsf{4. The Torsion-Induced Energy Flux}}}
Torsional Inflation is based on the simple idea that the extrinsic torsion of the distinguished spatial sections was, like the corresponding extrinsic curvature, very small at first: it becomes large only towards the end of Inflation, as a result of the decay of the intrinsic torsion, which is enforced ultimately by equation (\ref{C}).

This may help us to understand the purported observations \cite{kn:eoinlow} of anisotropies at relatively small scales. But can torsion also give a natural explanation of anisotropies on much larger scales?

We will argue now that it can, and the explanation arises from the unusual form taken by the \emph{Codazzi equation} in torsional geometry.

The Einstein-Cartan version of the familiar General-Relativistic field equation is
\begin{equation}\label{P}
\m{Ric}^*\,-\,{1\over 2}\,\m{Scal}^*\,g^*\,=\,{8\pi G\over c^4}\,P^*,
\end{equation}
where, as usual, the asterisk denotes a spacetime quantity, $\m{Ric}^*$ and $\m{Scal}^*$ are respectively the Ricci and scalar curvatures of the torsional connection, and $P^*$ is the stress-energy-momentum tensor.

Thus far we have only considered this equation as it governs the inflationary energy density. We now need to discuss the energy \emph{fluxes} quantified by $P^*.$

The energy flux within each spatial section $\Sigma$ is described by a one-form $\mathcal{F}$ given in terms of $P^*$ and the unit normal vector $\xi$ by
\begin{equation}\label{R}
\mathcal{F}(X)\,=\, - cP^*(\xi,\,X),
\end{equation}
for any $X$ tangent to $\Sigma$. Using equation (\ref{P}) we can express this as
\begin{equation}\label{S}
{8\pi G\over c^5}\,\mathcal{F}(X)\,=\,- \m{Ric}^*(\xi,\,X).
\end{equation}
Using the orthonormal basis defined by $\xi$, together with the spatial basis we used above when discussing equation (\ref{F}), we can write this (dispensing with the superscript notation, since the basis is orthonormal) as
\begin{equation}\label{T}
{8\pi G\over c^5}\,\mathcal{F}_j \,=\,- \sum_{i = 1}^3R^*_{i0ij},
\end{equation}
where all non-zero indices refer to the spatial basis.

Now $R^*_{\alpha\beta\gamma\delta}$ is antisymmetric in both the first and the second pairs of indices separately: this is true whenever the metric is compatible with the linear connection, as we always assume here. So we have
\begin{equation}\label{U}
{8\pi G\over c^5}\,\mathcal{F}_j \,=\,- \sum_{i = 1}^3R^*_{0iji}.
\end{equation}

To proceed, we need the torsional version of the usual (semi-Riemannian) Codazzi equation, the proof of which is given in the Appendix below. It is an equation for the normal component of the spacetime curvature $R^*$, when it is evaluated on a triple $X,\,Y,\,Z$ of tangent vectors to $\Sigma$: in signature $(-,\,+,\,+,\,+)$ it states that
\begin{equation}\label{V}
- g^*(\xi,\,R^*(X,\,Y)Z)\,=\,\left(\nabla_X h\right)(Y,\,Z)\,-\,\left(\nabla_Y h\right)(X,\,Z)\,+\,h(T(X,\,Y),\,Z),
\end{equation}
where $\xi$ is the unit normal field, $h$ is the second fundamental form, $\nabla$ is the purely spatial covariant derivative operator, and $T$ is the intrinsic torsion. Notice that, in the zero-torsion case, the right side of the Codazzi equation involves only the derivatives of $h$; but this is not necessarily so in the torsional case, where there is in general a purely algebraic contribution from the second fundamental form and the intrinsic torsion. This is the key point here.

The Codazzi equation allows us to express equation (\ref{U}) as
\begin{equation}\label{W}
{8\pi G\over c^5}\,\mathcal{F}_j \,=\, \sum_{i = 1}^3\left(\nabla_j h_{ii} \,-\, \nabla_i h_{ji}\,-\,\sum_{k = 1}^3h_{ki}\emph{\text{\v{T}}}_{kij}\right),
\end{equation}
where we remind the reader that $\emph{\text{\v{T}}}$ is the $(0,\,3)$ version of the intrinsic torsion.

We notice at once that the existence of a non-zero energy flux is primarily \emph{a question to be settled by the study of the extrinsic geometry of the distinguished spatial sections}. For the dominant object on the right side of this equation is the second fundamental form. It is true that the intrinsic geometry does put in an appearance, in the form of $\emph{\text{\v{T}}}$ and inside $\nabla$; but any attempt to use a symmetry argument here will certainly have to involve $h$.

In the zero-torsion case ($\emph{\text{\v{T}}} = 0$), when the spatial sections are isotropic around every point, we have $h = Hg/c,$ with zero spatial covariant derivative, and  so the energy fluxes are all zero: as usual in this case, the symmetries of the intrinsic geometry impose themselves on the extrinsic geometry, ruling out a distinguished direction that would be defined by an energy flux.

When the torsion is not zero, then of course the first term on the right side of equation (\ref{J}) still has zero spatial covariant derivative, and in fact so does, under our assumptions, the second term. To see this, note first that, as discussed earlier, the spatial connection coefficients vanish globally in the basis we are using here, so the spatial covariant derivative reduces in this basis to spatial directional derivatives of the only non-zero component of $\varkappa,$ namely the extrinsic torsion parameter $\kappa.$ As we are assuming that this quantity depends only on time, the final result is that the spatial covariant derivative of $h$ is zero, just as in the zero-torsion case. But this does not mean that the energy fluxes along the spatial axes have to be zero: instead they are given by
\begin{equation}\label{X}
\mathcal{F}_j \,=\, -\,{c^5\over 8\pi G}\sum_{i = 1}^3\sum_{k = 1}^3h_{ki}\emph{\text{\v{T}}}_{kij}.
\end{equation}
Using equation (\ref{J}) and the antisymmetry of $\emph{\text{\v{T}}}_{kij}$ in every pair of indices, we find that this is
\begin{equation}\label{Y}
\mathcal{F}_j \,=\, -\,{c^5\over 16\pi G}\sum_{i = 1}^3\sum_{k = 1}^3\varkappa_{ki}\emph{\text{\v{T}}}_{kij}.
\end{equation}
This equation, which derives from combining the field equation with the Codazzi relation, is \emph{the fundamental result for our work here}. It has three major consequences:

$\bullet$ When torsion is present, there will be in general an energy flux, even if nothing depends on spatial position. This is due to the final term on the right in equation (\ref{V}). The existence of such a flux is not unprecedented: such fluxes also arise in ordinary inflationary perturbation theory. The novelty here is that a specific direction will be singled out as the effect grows.

$\bullet$ That term is \emph{always} present in non-trivial cases, even if we replace the Einstein-Cartan theory by some much more complex theory allowing torsion to propagate: for it is part of the Codazzi relation, which is valid universally. A more complex theory will complicate the relation between spin and torsion, but it cannot of course affect the basic equations of geometry.

$\bullet$ The energy flux, being directional, can only exist because of the inherent anisotropy of extrinsic torsion, which is able to defy ``isotropisation''. However, the extrinsic torsion is small at this time. But this is where the appearance of the \emph{intrinsic} torsion in equation (\ref{Y}) is crucial: it is large at early times, and so, because its components occur as products with those of the extrinsic torsion, this serves to \emph{amplify} the anisotropy associated with the latter. The intrinsic torsion itself is isotropic, but the peculiar form taken by the torsional Codazzi relation allows it to give rise to anisotropy at early times in this indirect manner.

From equation (\ref{F}) we see that $\mathcal{F}_1 = \mathcal{F}_2 = 0,$ as one would expect from the rotational symmetry around the distinguished axis, whereas, in the distinguished direction, we have instead
\begin{equation}\label{Z}
\mathcal{F}_3 \,=\, -\,{c^5\over 8\pi G}\varkappa_{12}\emph{\text{\v{T}}}_{123},
\end{equation}
and this does not vanish if the spin density is non-zero.

In summary, there is inevitably an energy flux in the direction distinguished by the extrinsic torsion. This will leave its mark on the superhorizon physics at the relevant times, and that mark will inevitably be anisotropic. \emph{This is how the torsional anisotropy might be manifested physically on very large scales}\footnote{This anisotropy is distinct from any anisotropy in the Hubble tensor describing the expansion of the Universe at the \emph{present} time. That is a different matter altogether.}.

We note in passing that, recently, the possibility of cosmic scale heat flows has been explored in great detail in \cite{kn:tariq}. It would be very interesting to know whether such heat flows might be seeded by energy fluxes, of the kind we have shown to exist in torsional Inflation, in the early stages of Inflation.

From equations (\ref{D}), (\ref{F}), and (\ref{H}), we find that, when the spacetime metric is of the FRW form,
\begin{equation}\label{ALPHA}
|\mathcal{F}_3| \,=\, {c^5\over 2\pi G}{1\over a}\left({4\pi G \sigma \over c^3}\,-\,{1\over a}\right).
\end{equation}
Substituting from equation (\ref{O}) into this, we now have an exact expression for the energy flux induced by torsional anisotropy in the Einstein-Cartan theory\footnote{There is of course a momentum density associated with the energy flux. In General Relativity the two are always one and the same, up a factor of $c^2$, but in torsional theories that is not in general the case, since the Ricci tensor is not usually symmetric (as a bilinear form) in the presence of torsion. One can however use the torsional first Bianchi identity (\cite{kn:kobnom1}, Chapter III) to express $\m{Ric}^*(X,\,\xi)\,-\,\m{Ric}^*(\xi,\,X)$ as a combination of various components of the torsion and its derivatives; under the assumptions being made here (that the intrinsic and extrinsic torsions are the only non-zero spacetime torsion components, and that the intrinsic torsion is maximally symmetric, that is, fully antisymmetric) it turns out that this object vanishes identically. Thus, in our specific case, the momentum density is aligned with the distinguished direction and its magnitude is, up to the $c^2$ factor, the same as that of the energy flux.}.

We know (equation (\ref{O})) that $a(t)$ is a monotonically increasing unbounded function of time, and so the qualitative form of $|\mathcal{F}_3|$ as a function of time is easily deduced from equation (\ref{ALPHA}). It is zero at $t = t_0$ when Inflation begins, and then increases, reaching a maximum value
\begin{equation}\label{BETA}
|\mathcal{F}_3|_{\m{max}} \,=\, {2\pi G \sigma^2\over c} \,=\, {3c^5\over 8\pi G L^2_{\sigma}};
\end{equation}
this can be expressed in a clearer way by comparing the energy flux with the energy density, in the dimensionless form
\begin{equation}\label{BETABETA}
{|\mathcal{F}_3|_{\m{max}}\over \rho c} \,=\, {L_{\rho}^2\over L_{\sigma}^2}.
\end{equation}
After this, the flux decreases and approaches zero towards the end of Inflation.

Of course, this pattern ---$\,$ zero at first, a rise to a maximum, and a final decline ---$\,$ was to be expected: the effect of the growth of the extrinsic torsion is eventually halted and reversed by the decay of the intrinsic torsion. What is not obvious is this: \emph{when} is the flux maximal? In view of our discussion at the beginning of this Section, we need it to be focussed on the \emph{early} inflationary era. We now show that this is so.

The maximum flux is attained at a time $t_{\m{max}}$, which (by using equations (\ref{ALPHA}) and (\ref{O})) we can show to satisfy
\begin{equation}\label{BETH}
2\sqrt{1 + {L_{\sigma}^2\over L_{\rho}^2}}\,=\,\sinh(ct_{\m{max}}/\ell)\,-\,{\cosh(ct_{\m{max}}/\ell)\,-\,1\over \sqrt{1 + {L_{\sigma}^2\over L_{\rho}^2}}}.
\end{equation}
Similarly, for the time $t_0$ corresponding to the beginning of Inflation, equations (\ref{E}), (\ref{M}), and (\ref{O}) give us
\begin{equation}\label{GIMEL}
\sqrt{1 + {L_{\sigma}^2\over L_{\rho}^2}}\,=\,\sinh(ct_0/\ell)\,-\,{\cosh(ct_0/\ell)\,-\,1\over \sqrt{1 + {L_{\sigma}^2\over L_{\rho}^2}}}.
\end{equation}
As we regard $\ell$ as being fixed by the inflaton decay rate (as above), these equations define $t_{\m{max}}$ and $t_0$ as functions of $L_{\sigma}/L_{\rho};$ their difference, $\tau \equiv t_{\m{max}} - t_0,$ is \emph{the proper time from the start of Inflation to the time when the maximum energy flux is attained}, and this function of $L_{\sigma}/L_{\rho}$ is shown in Figure 1.
\begin{figure}[h]
\centering
\includegraphics[width=0.99\textwidth]{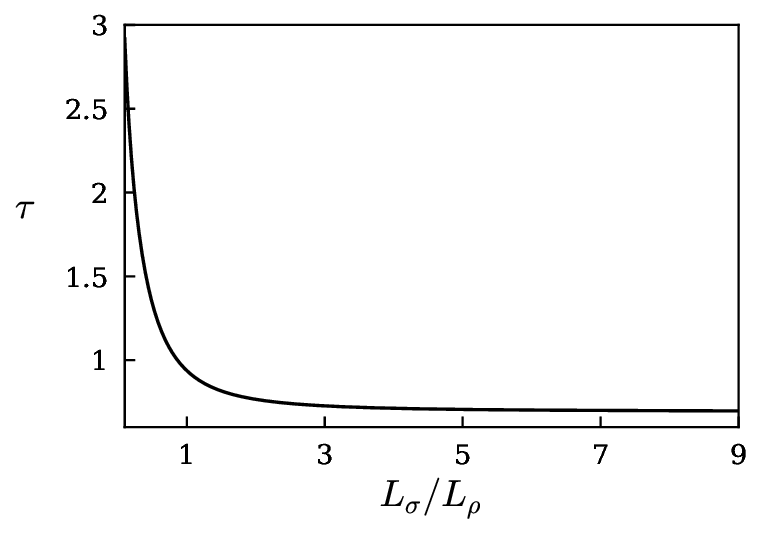}
\caption{Time after Inflation is initiated (in units of $\ell/c$, that is, ``late'' e-folds) at which the energy flux is maximal, as a function of the ratio of the length scales defined by the spin and energy densities.}
\end{figure}

We see that $\tau$ is monotonically decreasing, and is asymptotic to a non-zero value, which is approximately 0.69 in terms of late e-folds: \emph{the maximum energy flux can never occur any earlier in inflationary history than this}.

We wish to argue that it also cannot occur much later.

From Figure 1 we see that it \emph{could} occur much later if $L_{\sigma}/L_{\rho}$ can be very small. But we saw earlier that small values of $L_{\sigma}/L_{\rho}$ are forbidden by the requirements of BBN. In addition, equation (\ref{BETABETA}) implies that values of $L_{\sigma}/L_{\rho}$ below unity violate the \emph{Dominant Energy Condition} (DEC) \cite{kn:brand}, which is the standard way of enforcing causality in systems described by a stress-energy-momentum tensor. Admittedly, it has long been known \cite{kn:matt} that (non-minimally coupled) scalar fields can violate this condition (in a certain specific sense) without causing signal propagation outside the light cone; furthermore, it is not clear whether the theorem discussed in \cite{kn:brand} can be usefully extended to the torsional case. Whether the spin-bearing inflaton can be stable while violating the DEC remains to be seen.

Overall, it seems clear that values of $L_{\sigma}/L_{\rho}$ below unity should be viewed with a great deal of caution, and we think it reasonable to avoid them here. In that case, equation (\ref{BETABETA}) simply implies that the energy flux must be smaller than the energy density, which certainly seems reasonable.

One finds that $\tau(1) \approx 0.94$. Referring once again to Figure 1, we conclude that, approximately,
\begin{equation}\label{DALETH}
0.69\,\leq \,{c\tau \over \ell}\,\leq 0.94.
\end{equation}

We do not wish to insist that these precise numerical values must hold. The more important qualitative conclusion is that the anisotropic energy flux \emph{is concentrated at some time early in Inflation}, which is precisely what we were hoping to demonstrate.

We would also like to argue that there is no particular reason for $L_{\sigma}$ to differ very greatly from $L_{\rho}.$ If this is granted, then $L_{\sigma}/L_{\rho}$ might be a little larger than unity, so that the upper end of the range in (\ref{DALETH}) is favoured.

This assumption has two further important consequences.

First, from equation (\ref{BETABETA}), the energy flux is not very small relative to the energy density. This means that the anisotropic disturbance in the inflaton field is significant: it stands out from the usual random fluctuations to which that field is subject. We have a genuine seed for a substantial anisotropy at large scales.

Secondly, however, the fact that small values of $L_{\sigma}/L_{\rho}$ are forbidden means that this signal cannot be \emph{arbitrarily} large. In that sense, it is not surprising that this variety of cosmic anisotropy is not at all conspicuous in the observations. This is partly due to the rather striking linkage between BBN physics and large-scale structure we have argued for here.

To return to the timing of the flux: we estimate that the maximal flux occurred very early in inflationary history, no more than about one late e-fold after Inflation begins. Of course, the flux does not instantly ``switch off'', so the effect might be important for a few late e-folds after the maximum.

Unfortunately, the ``number of e-folds'' in Inflation, usually cited as \emph{at least} 50 to 60, is measured \emph{backwards} from its end \cite{kn:orazi0}. Thus, the statement that the energy flux is associated with the ``early'' stage of Inflation is difficult to interpret in terms of observations.

There has however long been theoretical interest in the possibility of ``just enough'' Inflation: that is, that the total number of e-folds might have been about equal to the minimum \cite{kn:guth,kn:remmen}. More recently, there has been much discussion \cite{kn:zal,kn:specogna,kn:kallosh} of hints in the observational data suggesting a very small positive value for the spatial curvature parameter\footnote{That would admittedly conflict with our discussion above, in which the (semi-Riemannian aspect of the) spatial geometry is spherical. We hope to return to this (apparent) contradiction elsewhere.}, and this too is suggestive of a relatively small total number of e-folds.

Let us assume that ``just enough'' Inflation is viable. In that context, the statement that a torsion-induced energy flux is important for a few e-folds after Inflation starts means that it leaves its mark \emph{on observations pertaining to large observable scales}, corresponding to perturbations that left the horizon at very early times. This might, as we hoped, allow torsion to explain anisotropy on the largest, as well as the smallest, cosmic scales.

Unfortunately, the present uncertainty as to the total duration of Inflation makes it difficult to be very specific as to what we should expect to see on large scales. A better understanding of ``just enough'' Inflation, particularly in the context of possible recent indications of a non-zero curvature parameter \cite{kn:zal,kn:specogna,kn:kallosh}, is needed here.

\addtocounter{section}{1}
\section* {\large{\textsf{5. Conclusion}}}
Recall (\cite{kn:kobnom2}, Chapter VIII) that the torsion tensor appears in the most fundamental equation governing manifolds with a linear connection, the equation of geodesic deviation:
\begin{equation}\label{DEV}
\nabla_u \nabla_u v\; +\;\,\nabla_u\left[T(v, u)\right] \;+\;R(v, u)u \;=\;0,
\end{equation}
where $\nabla$ is the connection, $R$ is its curvature, $T$ its torsion, $u$ is the tangent vector to a family of geodesics, and $v$ is the deviation vector.

Equation (\ref{DEV}) makes it clear that torsion stands on the same footing as curvature\footnote{Despite this, it is sometimes said that the inclusion of torsion complicates the question of interpreting gravitation in terms of effective field theory \cite{kn:crap}. As is pointed out in \cite{kn:donoghue}, however, it has never been shown conclusively that the EFT view of gravity can deal with fully general spacetimes even in standard General Relativity; so it is premature to criticise torsional theories on such grounds. (For an example of an effective field theory computation involving torsion, see \cite{kn:alm}.) It is probably best to regard this as a question to be settled ultimately by observations \cite{kn:squeeze} and to keep an open mind as to the outcome.}. In particular, there is nothing to be gained from pretending that torsion is ``just another matter field''. Where is the coupling constant in equation (\ref{DEV})?

Thus, the possibility of non-zero spacetime torsion certainly deserves to be taken very seriously. There are many ways in which this might be pursued. For example, the study of quantum-field-theoretic aspects of torsion might lead to predictions regarding potentially observable effects in gravitational waves \cite{kn:gravwav}. This is particularly appropriate, since in studying gravitational waves we are \emph{directly} observing non-trivial spacetime geometry (see \cite{kn:lake} for ideas regarding the direct measurability of spacetime geometry), and that might well include non-zero torsion \cite{kn:armin}. There are in fact many other systems in which torsional spacetime geometry might eventually be seen, almost as directly: see \cite{kn:capolupo} for a recent example.

It is nevertheless true that the physical effects of torsion have been difficult to discern. In many cases, this has a perfectly natural explanation. For example, if we assume as usual \cite{kn:reheat2} that the inflaton decays to particles after reheating, and that the spin density of those particles is proportional to their number density, then (since torsion appears quadratically in the Friedmann equation, equation (\ref{I})), one expects \cite{kn:lasenby,kn:pop} the torsional contribution to the ``energy'' density to decay according to $a^{-6},$ that is, far more rapidly than matter or radiation. One therefore expects it to be hard to find direct evidence for torsion after reheating. (On the other hand, if dark matter is actually a relic of the inflaton \cite{kn:sahni}, and the latter carries spin, then this conclusion might have to be corrected.)

But this argument does not apply to the inflationary era: and in fact this is precisely where torsion should appear most conspicuously. Yet that does not appear to be the case.

The present work is based on the observation that the extrinsic torsion of the spatial sections of the inflationary world is inherently anisotropic, and it grows monotonically with the inflationary expansion. This led us to expect to see some evidence of torsional anisotropy at small cosmic scales, corresponding to the late inflationary era.

Here we have argued that, unexpectedly, a quite different kind of torsional anisotropy might arise very early in that era, and help to explain putative observations at the \emph{largest} scales. Unfortunately, it may be that this rather remarkable effect, arising from the mathematics of torsion (in the form of the torsional Codazzi equation), arises too early for observation to be possible at the present time. Though that would be a disappointment, at least we would then have an example, in the inflationary context, where we can answer the most basic question regarding torsion: why is it not seen?

\addtocounter{section}{1}
\section* {\large{\textsf{APPENDIX: The Torsional Codazzi Relation}}}
In standard General-Relativistic FRW cosmology, the component of spacetime curvature evaluated on spacelike vectors, but normal to the spatial sections, vanishes. That is not always so in the torsional case, and, as we have seen, this has important consequences. To see how this works, we need the torsional Codazzi relation. Here we give a very short proof, following the methods of the standard reference, \cite{kn:kobnom2}, Chapter VII.

We begin with a reminder of the definition of the second fundamental form for a spacetime locally foliated by three-dimensional spatial sections: it is a $(1,\,1)$ tensor $\hat{h}$ on any spatial section $\Sigma,$ which measures the bending of $\Sigma$ into the spacetime by measuring how the unit normal vector field $\xi$ varies as one moves about in $\Sigma$:
\begin{equation}\label{GAMMA}
\hat{h}(X)\,=\,\nabla^*_X\xi.
\end{equation}
Here $X$ is a vector field tangential to $\Sigma$ and $\nabla^*$ is the covariant derivative of the spacetime connection. (To see that $\hat{h}(X)$ is tangential to $\Sigma,$ apply $\nabla^*_X$ to the equation $g^*(\xi,\,\xi) = - 1,$ and recall that we are assuming that $\nabla^*$ is a metric connection.)

If $X$ and $Y$ are tangential to $\Sigma$, then $\nabla^*_XY$ has a component tangential to $\Sigma,$ and this turns out to be $\nabla_XY,$ where $\nabla$ is the covariant derivative defined by the induced connection on $\Sigma.$ But $\nabla^*_XY$ also has a normal component, the magnitude of which is (using, for convenience, a dot to denote the ``scalar product'' defined by the spacetime metric $g^*$) $-\,\xi\,\cdot \,\nabla^*_XY = Y\,\cdot \,\nabla^*_X\xi = \hat{h}(X) \cdot Y = h(X,\,Y),$ where $h$ is the $(0,\,2)$ version of the second fundamental form. Consequently we have
\begin{equation}\label{DELTA}
\nabla^*_XY\,=\,\nabla_XY\,+\,h(X,\,Y)\xi.
\end{equation}
Therefore, if $Z$ too is tangential to $\Sigma,$ then
\begin{equation}\label{EPSILON}
\nabla^*_X\nabla^*_YZ\,=\,\nabla_X\nabla_YZ\,+\,h(X,\,\nabla_YZ)\xi\,+\,X(h(Y,\,Z))\xi \,+\, h(Y,\,Z)\hat{h}(X).
\end{equation}
The first term on the right is tangential to $\Sigma,$ and so too is the final term, as we saw above; and so the normal component, denoted by $\mathcal{N},$ is
\begin{equation}\label{ZETA}
\mathcal{N}\left(\nabla^*_X\nabla^*_YZ\right)\,=\,\left(h(X,\,\nabla_YZ)\,+\,X(h(Y,\,Z)\right)\xi.
\end{equation}
Similarly we have
\begin{equation}\label{ETA}
\mathcal{N}\left(\nabla^*_{[X,\,Y]}Z\right)\,=\,h([X,\,Y],\,Z)\xi,
\end{equation}
where the square brackets denote a commutator. The definition of torsion allows us to express this equation as
\begin{equation}\label{THETA}
\mathcal{N}\left(\nabla^*_{[X,\,Y]}Z\right)\,=\,h\left(\nabla_XY\,-\,\nabla_YX \,-\,T(X,\,Y),\,Z\right)\xi;
\end{equation}
this is the crucial step for us, since it shows where the intrinsic torsion mixes, in the final term, with the second fundamental form (and thus with the extrinsic torsion).

Combining equation (\ref{ZETA}) with the same equation with $X$ and $Y$ reversed, and with equation (\ref{THETA}), we find, using the definition of the spacetime curvature tensor and $(\nabla_Xh)(Y,\,Z) = X(h(Y,\,Z))\,-\,h(\nabla_XY,\,Z)\,-\,h(Y,\,\nabla_XZ),$ (and the same equation with $X$ and $Y$ reversed) that
\begin{equation}\label{IOTA}
\mathcal{N}\left(R^*(X,\,Y)Z\right)\,=\,\left[\nabla_Xh(Y,\,Z)\,-\,\nabla_Yh(X,\,Z) \,+\,h\left(T(X,\,Y),\,Z\right)\right]\xi;
\end{equation}
and this is exactly the Codazzi relation, equation (\ref{V}).

\end{document}